# Clifford algebra approach of 3D Ising model

Zhidong Zhang

Shenyang National Laboratory for Materials Science, Institute of Metal Research, Chinese Academy of Sciences, 72 Wenhua Road, Shenyang, 110016, P.R. China

Corresponding author's e-mail: zdzhang@imr.ac.cn

Osamu Suzuki

Department of Computer and System Analysis, College of Humanities and Sciences, Nihon University, Sakurajosui 3-25-40, 156-8550 Setagaya-ku, Tokyo, Japan

E-mail: osuzuki1944butterfly@gmail.com

Norman H. March

Oxford University, Oxford, England

E-mail address: tony@lonsdaleconsulting.co.uk.

We develop a Clifford algebra approach for 3D Ising model. We first note the main difficulties of the problem for solving exactly the model and then emphasize two important principles (i.e., Symmetry Principle and Largest Eigenvalue Principle) that will be used for guiding the path to the desired solution. By utilizing some mathematical facts of the direct product of matrices and their trace, we expand the dimension of the transfer matrices **V** of the 3D Ising system by adding unit matrices **I** (with compensation of a factor) and adjusting their sequence, which do not change the trace of the transfer matrices **V** (Theorem I: Trace Invariance Theorem). The transfer matrices **V** are re-written in terms of the direct product of sub-transfer-matrices $Sub(V^{(\delta)}) = [I \otimes I \otimes ... \otimes I \otimes V^{(\delta)} \otimes I \otimes ... \otimes I]$, where each $V^{(\delta)}$ stands for the contribution of a plane of the 3D Ising lattice and interactions with its neighboring plane. The sub-transfer-matrices $V^{(\delta)}$ are isolated by a large number of the unit matrices, which allows us to perform a linearization process on $V^{(\delta)}$ (Theorem II: Linearization Theorem). It is found that locally for each site j, the internal factor $W_j$ in the transfer matrices can be treated as a boundary factor, which can be dealt with by a procedure similar to the Onsager-Kaufman approach for the boundary factor U in the 2D Ising model. This linearization process splits each sub-transfer matrix into $2^n$ sub-spaces (and the whole system into $2^{nl}$ sub-spaces). Furthermore, a local transformation is employed on each of the sub-transfer matrices (Theorem III: Local Transformation Theorem). The local transformation trivializes the non-trivial topological structure, while it generalizes the topological phases on the eigenvectors. This is induced by a gauge transformation in the Ising gauge lattice that is dual to the original 3D Ising model. The non-commutation of operators during the processes of linearization and local transformation can be dealt with to be commutative in the framework of the Jordan-von Neumann-Wigner procedure, in which the multiplication $A \circ B = \frac{1}{2}(AB + BA)$ in Jordan algebras is applied instead of the usual matrix multiplication AB (Theorem IV: Commutation Theorem).. This can be realized by time-averaging t systems of the 3D Ising models with time evaluation. In order to determine the rotation angle for the local transformation, the star-triangle relationship of the 3D Ising model is employed for Curie temperature, which is the solution of generalized Yang-Baxter equations in the continuous limit. Finally, the topological phases generated on the eigenvectors are determined, based on the relation with the Ising gauge lattice theory. ..

## 1. Introduction to 3D Ising model

The study on the Ising model has a long history in fields of both physics and mathematics [9]. The Ising model is based upon the view that ferromagnetism is caused by an interaction between the spins of certain electrons in the atoms making up a crystal [28]. A spin coordinate σ is associated with each lattice point of the crystal, which is considered as a scalar quantity. It can achieve either of two values σ = ±1, corresponding to either spin-up or spin down state. Usually, only the interaction energy between two spins located at the nearest neighbors of the lattice points is taken into account [25,28]. The thermodynamic and magnetic properties of a crystal which contains N lattice points can be determined from the partition function. The Ising model describes the transition from a paramagnetic to a ferromagnetic phase in a magnetic lattice and becomes a paradigm for many different systems, including antiferromagnets, lattice gases, and large biological molecules, etc. [25]. In the case of two-dimensional (2D) Ising model, no non-trivial topological structures exist [29]. The existence of topological effect in three-dimensional (3D) Ising model has been pointed out in [28] and proofed rigorously in [10]. As pointed out in Newell and Montroll's paper [28], the combinatorial method of counting the closed graph for the 2D Ising model cannot be generated in any obvious way to the 3D problem (see page 366 in [28]). For the 3D Ising model, one encounters polygons with knots (see page 367 in [28]). The peculiar topological property is that a polygon in three dimensions does not divide the space into an "inside and outside" (see page 367 in [28]). Indeed the 3D lattices are inherently nonplanar and any nonplanar graph throws up a barrier of computational intractability. Istrail showed that the essential ingredient in the

NP-completeness of the Ising model is nonplanarity [10]. Every nonplanar lattice in two or three dimensions contains a subdivision of an infinite graph he called the "basic Kuratowskian" [10]. On the observation of the formula of the partition functions of 3D Ising model, the first author (ZDZ) conjectured that the non-trivial topological structures of the 3D Ising model can be trivialized in a higher dimensional space and the 3D Ising model can be realized as the free statistic model on them with weight factors (i.e., topological/geometrical phases) on eigenvectors [44,45]. Based on the two conjectures proposed in [44], a putative exact solution is derived for the 3D Ising model [44].

There have been several rounds of comments/responses between some statistical physicists and Zhang [30-32,39,40,46,47]. We do not want to repeat Zhang's responses published already [45-50]. Just mention briefly: The error for the Jordan-Wigner transform in Zhang's original paper [44] is not a problem, which has been corrected in ref. [45] and also the present manuscript, Zhang's conjectures can start from the corrected formula. The topological effects indeed exist as clearly seen from the corrected Jordan-Wigner transform. Because of neglecting the topological effect, the so-called exact and rigorous approaches of 1960s and later are not exact and rigorous. The so-called exact low-temperature expansion is evidently divergent, which indicates that this expansion approach is not exact, not only for its high-order terms, but also for the first term, since the approach itself is questionable. The Lee-Yang theorems for phase transition offer a possibility of a phase transition at infinite temperature [23,41], which provides a possibility of multi-valued functions for high-temperature expansions. It should be noticed that the convergences of βf and f are different at/near infinite temperature.

In this work, we develop a Clifford algebra approach for the 3D Ising model, which gives a positive answer to the Zhang's two conjectures. .

At first, we recall some basic facts on the 3D Ising model. We consider the Ising Hamiltonian on an orthorhombic lattice in 3D Euclidean space [44,45], with up-spin or down-spin at each lattice point:

$$H = -J\sum_{\tau=1}^{n}\sum_{\rho=1}^{m}\sum_{\delta=1}^{l} s_{\rho,\delta}^{(\tau)} s_{\rho,\delta}^{(\tau+1)} - J'\sum_{\tau=1}^{n}\sum_{\rho=1}^{m}\sum_{\delta=1}^{l} s_{\rho,\delta}^{(\tau)} s_{\rho+1,\delta}^{(\tau)} - J''\sum_{\tau=1}^{n}\sum_{\rho=1}^{m}\sum_{\delta=1}^{l} s_{\rho,\delta}^{(\tau)} s_{\rho,\delta+1}^{(\tau)}. \qquad (1)$$

Here only the nearest interaction between spins at each lattice point is considered. In Appendix A, we show the process that derives the three transfer matrices $\mathbf{V_1}$, $\mathbf{V_2}$ and $\mathbf{V_3}$ from the Hamiltonian (1). The partition function of the 3D Ising model can be given as follows [45]:

$$Z = (2\sinh 2K)^{\frac{m\cdot n\cdot l}{2}} \cdot trace(V_3 V_2 V_1)^m \equiv (2\sinh 2K)^{\frac{m\cdot n\cdot l}{2}} \cdot \sum_{i=1}^{2^{n\cdot l}} \lambda_i^m \qquad (2)$$

$$V_3 = \prod_{j=1}^{nl} \exp\{iK''\Gamma_{2j}\left[\prod_{k=j+1}^{j+n-1} i\Gamma_{2k-1}\Gamma_{2k}\right]\Gamma_{2j+2n-1}\}; \qquad (3)$$

$$V_2 = \prod_{j=1}^{nl} \exp\{iK'\Gamma_{2j}\Gamma_{2j+1}\}; \qquad (4)$$

$$V_1 = \prod_{j=1}^{nl} \exp\{iK*\cdot\Gamma_{2j-1}\Gamma_{2j}\}. \qquad (5)$$

Here we introduce the following generators of Clifford algebra of 3D Ising model:

$$\Gamma_{2j-1} = C\otimes C\otimes......\otimes C\otimes s'\otimes I \otimes...\otimes I \quad (j\ times\ C) \qquad (6)$$

$$\Gamma_{2j} = C\otimes C\otimes......\otimes C\otimes (-is'')\otimes I \otimes...\otimes I \quad (j\ times\ C) \qquad (7)$$

The Clifford algebra is generated by Pauli matrices. The problem becomes to how to deal with the trace of the transfer matrices $\mathbf{V} = \mathbf{V_3}\mathbf{V_2}\mathbf{V_1}$. Notice that the transfer matrix $\mathbf{V_1}$ or $\mathbf{V_2}$ with the product of two $\Gamma$ matrices has trivial topologic structure, which can be represented as Lie group for a rotation. The transfer matrix $\mathbf{V_3}$ is the root of difficulties to hinder the application of the Onsager-Kaufman approach [18,29] to 3D Ising model, due to the non-linear terms of $\Gamma$-matrices. In the transfer matrix $\mathbf{V_3}$, the internal factor is denoted as $W_j = \left[ \prod_{k=j+1}^{j+n-1} i\Gamma_{2k-1}\Gamma_{2k} \right]$. The main difficulties arisen by the transfer matrix $\mathbf{V_3}$ are described briefly as follows:

**1) Non-local behavior:** With the language of the fermionic ($\Gamma$) variables, it is clear that there are non-trivial topological effects in the 3D Ising model, which are non-local behavior. The topological effects can be seen also in the language of the spin ($\sigma$) variables although all expressions are local. The third transfer matrix $\mathbf{V_3}$ must follow the sequence of the spin ($\sigma$) variables arranged already in the first two transfer matrices $\mathbf{V_1}$ and $\mathbf{V_2}$. According to this sequence, although the interaction between a spin and one of its neighboring spins along the third dimension is the nearest neighboring one, its effect is correlated with the states of n other spins in the plane. As revealed in Zhang's recent paper [49], the global effects indeed exist in the 3D Ising system, in which all the spins are entangled. It was understood that the different topological states (e.g. knots/links) formed by up or down spins also contribute to the partition function and hence the free energy and other physical properties of the 3D Ising system [45]. It is clear that any approaches based on only local environments, such as conventional low-temperature expansions, conventional high-temperature expansions, Monte Carlo simulations, Renormalization Group, etc, are not exact

[44,45].

2) **Non-Gaussian**: Each fermionic Gaussian corresponds to a rotation, which is the spinor representation of the rotation group, i.e., the Lie group. The spinor representation of the Lie algebra corresponding to rotations consists of quadratic expressions in the Γ-matrices only. Expressions that are of higher degree than quadratic in Γ-matrices (like the log of $V_3$), are not elements of the Lie algebra [32]. Thus an element of exponential factors with higher degree of Γ-matrices (so-called non-linear terms) is not an element of the Lie group. The problem becomes to how to linearize the high-order terms to be an element of the Lie group, which is the spinor representation of the rotation.

3) **Non-commutative:** In two dimensions, we can apply the methods of Kaufman, since there is only one fermion string, the product of all Γ-matrices, which commutes with every part of the transfer matrix. Then we can project to the two subspaces with eigenvalues +1 and -1 of the string [18]. In the 3D Ising model, the transfer matrix $V_3$ has many factors in the product over j, each of them has an "internal factor" $W_j$, thus there are many internal factors that must be diagonalized simultaneously and commute with the whole $V_3$. Even though each internal factor $W_j$ commutes with both $\Gamma_{2j}$ and $\Gamma_{2j+2n-1}$, these two Γ-matrices do not commute, but anticommute [18]. Therefore, it seems that there is no way to diagonalize the internal factor $W_j$, $\Gamma_{2j}$ and $\Gamma_{2j+2n-1}$ simultaneously. Furthermore, even with one j, it mess up $V_2$ and $V_1$ when one attempts to use the non-linear transformation of the Γ-matrices. Even though the factors in the j-th factor of $V_3$ commute, the internal factors $W_j$ for different j do not commute with the other factors in $V_3$. When we diagonalize all the internal factors for different j simultaneously, the different $\Gamma_{2j}\Gamma_{2j+2n-1}$ are not simultaneously block diagonal in such a basis. When we may have some tricks, for instance, to redefine the two Γ-matrices, to work for a single j, this cannot be done for all j's at the same time.

Before performing our process to solve the problem, we emphasize two principles, which are very important for guiding the path to the desired solution.

**Symmetry Principle:** *The exact solution of the 3D Ising model is constructed by the basic forms that satisfy the symmetry of the system.*

At first, the symmetry of the 3D Ising model requires that the transfer matrices $\mathbf{V_2}$ and $\mathbf{V_3}$ have the same contribution to the exact solution. The exact solution must keep invariant no matter which crystallographic direction is chosen as the second or third dimension. The internal factor (i.e., non-linear terms) will appear in the transfer matrix $\mathbf{V_2}$ if we choose the $K''$ direction as the second crystallographic axis. Actually, if one transforms the transfer matrix $\mathbf{V_3}$ to be linear by a non-linear transformation, $\begin{cases} \Gamma'_{2j} = \Gamma_{2j} \\ \Gamma'_{2j+1} = \left[ \prod_{k=j+1}^{j+n-1} i\Gamma_{2k-1}\Gamma_{2k} \right] \Gamma_{2j+2n-1} \end{cases}$, the transfer matrix $\mathbf{V_2}$ will become a non-linear one by such a transformation. Although the high-order terms in the transfer matrices cannot be removed by this transformation, we can figure out that their contribution to the exact solution is merely a scaling factor, as seen clearly from the fact that the linear terms can appear in either $\mathbf{V_2}$ or $\mathbf{V_3}$. This symmetry indicates clearly that the exact solution of the 3D Ising model should be in the basic forms consisting of $\sinh 2(K'+\lambda K'')$ and $\cosh 2(K'+\lambda K'')$ [in consideration of the symmetry, it can be written also in terms of $\sinh 2(\lambda' K'+K'')$ and $\cosh 2(\lambda' K'+K'')$]. These forms satisfy the symmetric requirement under the non-linear transformation. Furthermore, when either $K'$ or $K''$ is zero, the exact solution is reduced to that of the 2D Ising model.

**Largest Eigenvalue Principle:** *Only does the largest eigenvalue contributes dominantly to the partition function of the 3D Ising model in the thermodynamic limit.*

This principle is very evident. Since the partition function Z is proportional to the sum of $\lambda_i^m$ (see Eq. (2)), only the largest eigenvaule of the transfer matrices **V** becomes dominant in the thermodynamic limit (m → ∞). Onsager [29] and Kaufman [18] discussed the largest eigenvaule and found that the exact partition function of the 2D Ising model does not differ much from the approximation result, in which all the eigenvalues but the largest are neglected. For the 3D Ising model, the same principle is valid, and we not only follow the Onsager-Kaufman procedure using the largest eigenvaule to find the formula similar to that of the 2D Ising model [18,29], but also utilize this principle to find the desired solution after realizing the local transformation (see Section 4). The linearization process and the local transformation, jointed with the Largest Eigenvalue Principle, state that the non-linear terms in the transfer matrices **V** merely contribute a scaling factor to the exact solution of the 3D Ising model, while generating topological phases. Note that the former is consistence with the result obtained above from Symmetry Principle.

**2. Trace invariance theorem for 3D Ising model**

From Eqs. (1) – (5) and (A1)-(A11), we have shown how to construct the partition function from the Hamiltonian of the 3D Ising model. It is illustrated clearly in Eq. (A4) that for the transfer matrix $\mathbf{V_1}$, there is equality between the direct product of the exponential factors of Pauli matrices C and the product of the exponential factors of matrices $C_j$. From Eqs. (A8) and (A11), it is seen that a similar equality is held for $s'_j s'_{j+1}$ of the transfer matrix $V_2$ (or $s'_j s'_{j+n}$ for the $V_3$) . Clearly, a situation for the transfer matrix $V_3$ is similar to other two transfer matrices $V_1$ and $V_2$: at the first beginning, they start from a formula for the direct product of the exponential factors of Pauli matrices. In this work, we will show that by utilizing some mathematical facts of the direct product of matrices and the trace, we can overcome the difficulties listed above so that we can deal with the terms for each row of the 3D Ising lattice, separately.

We have the following theorem.

**Theorem I (Trace Invariance Theorem)**

*The partition function of the 3D Ising model is changed a factor of $2^k$ by adding k terms of unit matrices in the direct product of the original transfer matrices. This factor of $2^k$ can be compensated by division to keep the invariance of the trace. Adjusting the sequence of the unit matrices with other matrices separates the exponential factors for different row (each row consists of n sites) in the transfer matrices. The exponential factors for each row are isolated by the identify matrices, so that they can be treated as sub-matrices of the transfer matrices to be dealt with separately.*

**Proof:**

It is noticed that we can obtain the transfer matrices $V_1$, $V_2$ and $V_3$ from the direct product of Pauli matrices. Therefore, we shall utilize the properties of the direct product of matrices, which are represented in Appendix B. By use of the generators of the Clifford algebra of the 3D Ising model, we can prove Theorem I by the following two steps.

The first step is to add the direct product of k (say, k = nl(o-1)) terms of unit matrices **I** to each transfer matrix **V₁**, **V₂** and **V₃**, and to divide the trace of the original transfer matrices **V** by $2^k$. Because $(I \otimes A) \cdot (I \otimes B) \cdot (I \otimes C) = I \otimes (ABC)$, adding one unit matrix to each transfer matrix **V₁**, **V₂** and **V₃** is equivalent to adding one unit matrix to the transfer matrix **V** = **V₃V₂V₁**. Since adding one unit matrix to **V** doubles its trance, this factor of $2^k$ for k terms of unit matrices **I** is used to compensate in order to keep the invariance of the trace. Meanwhile, the eigenvectors of the 3D Ising model are expanded to be fit with the (3+1)-dimensions. In this way, we represent the transfer matrix **V** and also the eigenvectors $\Psi$ in the (3+1)-dimensional framework. We can apply this algebra approach on the trace of the direct product of matrices (the left hand of the equations (A4), (A8) and (A11)) by adding the unit matrices, which does not change the trace of the transfer matrices **V**, according to the following equalities:

$$Trace(V) = Trace(V_3 V_2 V_1) = \frac{1}{2^k} Trace[I \otimes I \otimes ... \otimes I \otimes V_3 V_2 V_1] \qquad (8)$$

Furthermore, we have the following equalities:

$$Trace(V) = Trace[V^{(1)} \otimes V^{(2)} \otimes ... \otimes V^{(l)}] = \frac{1}{2^k} Trace[I \otimes I \otimes ... \otimes I \otimes V^{(1)} \otimes V^{(2)} \otimes ... \otimes V^{(l)}]$$

$$(9)$$

Here $V^{(1)}$, $V^{(2)}$, …, or $V^{(l)}$ denotes a small transfer matrix $V^{(\delta)}$ ($\delta$ = 1,2,…,l) in the transfer matrices **V**, which represents the contribution of a row of n sites of the 3D Ising lattice (if one considered the periodic condition used already for m lines (see Eq. (2)), it actually represents the contribution of nm sites in a plane) together with the interactions to its neighboring row (or plane if considered m lines). It equals to the product of the direct product of n terms of a matrices in **V₁**, the product of two matrices (each of them) consisting of the direct product of n/2 terms of b matrices in **V₂**, and the product of n terms of c matrices in **V₃** (see Eqs. (A5), (A9) and (A10) for details), $V^{(\delta)}$ can be represented in the following form: $V^{(\delta)} = V_3^{(\delta)} V_2^{(\delta)} V_1^{(\delta)}$ where $V_3^{(\delta)}$, $V_2^{(\delta)}$ and $V_1^{(\delta)}$ have the same feature as the transfer matrices **V₃**, **V₂** and **V₁** represented in Eqs. (3) – (5), but here only with n terms in each product.

The second step is to separate the small transfer matrices $V^{(1)}$, $V^{(2)}$, …, and $V^{(l)}$ from each other by adjusting the sequence of the unit matrices, with respect to these matrices. This procedure will decouple every row from each other, each of them corresponds to the products of n exponential factors in the transfer matrices **V₁**, **V₂** and **V₃** (see Eqs. (A5), (A9) and (A10), or Eqs. (3)-(5)) The product of these n exponential factors are surrounded by a large number (say n(o-1)) of the unit matrices. We have the following equalities:

$$Trace(V) = \frac{1}{2^k} Trace\left[I \otimes ... \otimes I \otimes V^{(1)} \otimes I \otimes ... \otimes I \otimes V^{(2)} \otimes I \otimes ...... \otimes I \otimes V^{(l)} \otimes I \otimes ... \otimes I\right]$$
$$= \frac{1}{2^k} Trace\left[Sub(V^{(1)}) \otimes Sub(V^{(2)}) \otimes ...... \otimes Sub(V^{(l)})\right]$$

(10)

Here $Sub(V^{(\delta)}) = [I \otimes ... \otimes I \otimes V^{(\delta)} \otimes I \otimes ... \otimes I]$ (δ = 1,2,…,l) denotes a sub-transfer matrix consisting of $V^{(1)}$, $V^{(2)}$, …, or $V^{(l)}$ with a large number (= n(o-1)) of unit matrices **I**. Since in anyone of these sub-transfer matrices, the exponential factors can be surrounded by a large amount of the unit matrices, they could temporally "forget" the actions of other exponential factors in the product. Due to $A \otimes B \otimes C = A \otimes (B \otimes C)$, we can first calculate the direct product in each sub-transfer matrix $Sub(V^{(\delta)})$, and then the direct product of all the sub-transfer matrices $Sub(V^{(\delta)})$. We can separately deal with the main elements (i.e., the exponential

factors) in these sub-transfer matrices

∎

By adding the identify matrices and keeping the trace invariance (Theorem I), we actually perform a decoupling process, which isolates every sub-transfer matrix $Sub(V^{(\delta)})$ in the transfer matrices from each other, so that we can attempt to diagonalize separately the sub-transfer matrices $Sub(V^{(\delta)})$ for each a row (or a plane as m lines considered already by the periodic boundary condition in Eq. (2)) interacting with its neighboring row (or plane). This means that we can deal with each sub-transfer matrix in the quasi-2D limit. The Theorem I verifies the following facts (as proposed in [44]): 1) The 3D Ising model can be described with the framework of (3+1)-dimensions. 2) The desired solution of the 3D Ising model possesses the feature of the Onsager's exact solution for the 2D Ising model.

**Remark:**

The Theorem I provides a procedure to first expand the dimension of the 3D Ising model to be (3 + 1), and then reduce the 3D Ising model, in order to separate Ising models in 2D manifolds (planes), where it is exactly solvable. Meanwhile, this procedure maintains the physical properties (such as the trace of the transfer matrices, the partition function, the free energy and the thermodynamic properties) of the system unchanged. This procedure allows us to apply some of Onsager-Kaufman procedures developed for the 2D Ising model.    .

In the next section, we will realize a linearization process for each sub-transfer matrix $Sub(V^{(\delta)})$. Then we can work all the way for all δ rows (see Section 3). Meanwhile, we can perform a local transformation on the sub-transfer matrices

$Sub(V^{(\delta)})$, which trivializes the non-trivial topological structure, while generalizing topological phases on eigenvectors (see Section 4). We will deal with the non-commutation of operators during the processes of linearization and local transformation (see Section 5). We derive the rotation angle of the local transformation from the symmetry of the system and the duality relationship (see Section 6). Finally, the topological phases on the eigenvectors are determined in Section 7.

**3. Linearization theorem for 3D Ising model**

We can prove the following theorem:

**Theorem II (Linearization Theorem)**

*A linearization process can be performed on the sub-transfer matrices $Sub(V^{(\delta)})$ of the 3D Ising model for each row δ. Locally, the non-linear terms in the sub-transfer matrices $Sub(V^{(\delta)})$ for each row δ can be linearized.*

**Proof**

One of the key steps to solve our problem is how to linearize the non-linear terms in the transfer matrices **V**. Due to the non-commutative properties of the Γ-matrices for different j, it is difficult to linearize these non-linear terms of Γ-matrices as a whole. Fortunately, following Theorem I, one finds that the δ-th row of the 3D Ising model has been decoupled from other rows. Then we can attempt to perform a linearization process independently for the sub-transfer matrices $Sub(V^{(\delta)})$ on the δ-th row of the 3D Ising system. In consideration of the periodic boundary condition used already along the line (m sites) for the transfer matrices **V**, a sub-transfer matrix $Sub(V^{(\delta)})$ represents a 2D Ising lattice (nm sites) interacting with one of its neighboring planes. We point out here a fact that the internal factor $W_j$ for the 3D Ising model has the form very much similar to the boundary factor U for the 2D Ising model. Therefore, we can develop a procedure generalized from Kaufman's approach for the boundary factor U for 2D Ising model, to study the effect of the internal factor $W_j$ for the 3D Ising model.

Kaufman [18] made the remarkable observation that one can decompose the transfer matrices **V** of the 2D Ising model as the product of factors like $e^{\theta \Gamma \Gamma/2}$, which is interpreted as a 2D rotation with rotation angle θ in the direct product space. The transfer matrices **V** of the 2D Ising model are decomposed into 2 pieces of subspaces in accordance with factors $\frac{1}{2}(I \pm U)$.[18] where U=$C_1 C_2....C_n$ = $\Gamma_1 \Gamma_2 \Gamma_3...\Gamma_{2n}$ is the boundary factor (see pages 1234 and 1237 of [18] for definition). One has the following relation $s_1 s_n = i P_1 Q_n U = i \Gamma_1 \Gamma_{2n} U$. We recall briefly the Kaufman's procedure for dealing with the boundary factor U of the 2D Ising model. According to Eq. (35) of [18], we have $V = \frac{1}{2}(I+U) V + \frac{1}{2}(I-U) V = \frac{1}{2}(I+U) V^+ + \frac{1}{2}(I-U) V^-$. Namely, in 2D Ising with periodic boundary conditions, the projection operators with U split the space into half the states of one such free-fermion problem with periodic boundary conditions in the fermions and half of the states of another free-fermion problem with antiperiodic boundary conditions [34]. Then the eigenvalues are selected in the two sub-spaces (for details, referred to pages 1238 and 1239 of ref. [18]).

For 3D Ising model, besides the boundary factors $U'$ and $U''$ (which are the same as that U for 2D Ising model), it is important to deal with the internal factor

$W_j = \left[ \prod_{k=j+1}^{j+n-1} i\Gamma_{2k-1}\Gamma_{2k} \right]$ in the transfer matrix $V_3$. We have the following relation $s_j s_{j+n} = i\Gamma_{2j}\Gamma_{2j+2n-1}W_j$. Clearly, the internal factors $W_j$ have the same character as the boundary factors, which can be linearized by use of the Kaufman approach for the boundary factor U of the 2D Ising model. However, for the sub-transfer matrices $Sub(V^{(\delta)})$, there are n terms with the internal factors $W_j$. Following the Kaufman's results (see Eq. (35) of her paper [18]), we have the following relation for dealing with one of these internal factors.

$$V^{(\delta)} = \tfrac{1}{2}(I+W_j) \ V^{(\delta)} + \tfrac{1}{2}(I-W_j) \ V^{(\delta)} = \tfrac{1}{2}(I+W_j) \ V^{(\delta)^+} + \tfrac{1}{2}(I-W_j) \ V^{(\delta)^-}. \quad (11)$$

Then we can split the sub-transfer matrix $V^{(\delta)}$ into two subspaces with the upper and lower squares of $^+V^{(\delta)}$. In the 2D Ising model, we can apply the methods of Kaufman, since there is only one fermion string, the product of all Γ-matrices, which commutes with every part of the transfer matrix, so that we can project to the two subspaces with eigenvalues +1 and -1 of the string [18]. In the 3D Ising model, even after the decouple process (Theorem I), we have to treat n terms with internal factors $W_j$. When we split the space of the sub-transfer matrices $V^{(\delta)}$ by one of the project operator $W_j$,, we cannot guarantee to keep the eigenvalues +1 and -1 for the subspaces. This problem should be fixed by a local transformation (see Theorem III). The sub-transfer matrices $V^{(\delta)+}$ and $V^{(\delta)-}$ in two subspaces of the j-th site may be written as:

$$V^{(\delta)\pm} = \tfrac{1}{2}(1 \pm W_j) \circ \prod_{j=1}^{n} \circ \exp\{iK*\cdot\Gamma_{2j-1}\Gamma_{2j}\} \prod_{j=1}^{n} \circ \exp\{iK'\Gamma_{2j}\Gamma_{2j+1}\}$$
$$\circ \exp\{\pm iK''\Gamma_{2j}\Gamma_{2j+2n-1}\} \prod_{j=1}^{n-1} \circ \exp\{iK''\Gamma_{2j}W_j\Gamma_{2j+2n-1}\} \quad (12)$$

In Eq. (12), multiplication $A \circ B = \dfrac{1}{2}(AB + BA)$ for two operators **A** and **B** are valid in Jordan algebra [12,13]. We aware that to obtain the formula (12) and also the following ones in next section, we must face the non-commutation of operators. This problem can be dealt with to be commutative in the framework of the Jordan-von Neumann-Wigner procedure [14] using Jordan algebras [12,13] and also performing a time average of many 3D Ising systems (see Theorem IV: Commutation Theorem).

We have to perform the same procedure to split $V^{(\delta)+}$ and $V^{(\delta)-}$ into four subspaces $V^{(\delta)++}$, $V^{(\delta)+-}$, $V^{(\delta)-+}$ and $V^{(\delta)--}$. It also generates the factors of $\frac{1}{2}(1 \pm W_j)$. After going on the same procedure to deal with all the n terms of the internal factors $W_j$ in a sub-transfer matrix $V^{(\delta)}$, we obtain $2^l \times (2 \times 2^n)$ pieces of sub-spaces (here the internal factors $W_j$ are responsible for $2^n$, while others are due to the usual boundary factors $U'$ and $U''$). There are the products of n terms of $\frac{1}{2}(1 \pm W_j)$ with the combination of $\pm$ signs for different piece [24]. For instance, one of these subspaces can be written as:

$$V^{(\delta)+\ldots+} = \frac{1}{2^n}\prod_{j=1}^{n} \circ (1+W_j) \prod_{j=1}^{n} \circ \exp\{iK*\cdot\Gamma_{2j-1}\Gamma_{2j}\}$$
$$\prod_{j=1}^{n} \circ \exp\{iK'\Gamma_{2j}\Gamma_{2j+1}\}\prod_{j=1}^{n} \circ \exp\{iK''\Gamma_{2j}\Gamma_{2j+2n-1}\} \quad (13)$$

Once again, the non-commutation of operators is solvable by the Jordan-von Neumann-Wigner procedure [14] with Jordan algebras multiplication $A \circ B = \frac{1}{2}(AB + BA)$ (see Theorem IV: Commutation Theorem). It is seen that the sub-transfer matrices in this sub-space are linear, and it is true for all the $2^n$ sub-spaces. For other sub-spaces, some signs before $W_j$ and/or $K''$ may be negative. The same procedure can be done for all the sub-transfer matrices $V^{(\delta)}$ ($\delta$= 1,2,…,l), which results in many eigenvalues in many pieces of subspaces. Therefore, the transfer matrices **V** of the 3D Ising model can be linearized and decomposed into $2^l \times (2 \times 2^{nl})$ pieces [24], in accordance with the projection operators $\frac{1}{2}(I \pm U')$, $\frac{1}{2}(I \pm U'')$ and $\frac{1}{2}(I \pm W_j)$, where $U'$ and $U''$ are the boundary factors and $W_j$ is the internal factor.. As usual, we can neglect the boundary factors in the thermodynamic limit, so the number of the pieces of subgroups may be reduced to $2^{nl}$. Of course, only one piece will produce the largest eigenvalue, which dominates the partition function in the thermodynamic limit, according with the Largest Eigenvalue Principle.

∎

**Remark:**

The Theorem II provides a procedure to deal with the 3D Ising model in 2D manifolds (planes). The cost of the linearization is the split of spaces. In this way, the non-linear problem in the whole 3D Ising system is transformed to the linear problems in the sub-spaces, while the non-linear character of the 3D Ising model is still kept from a global view. This procedure allows us to perform a series of plane rotations in local coordinates (sub-spaces) (see Theorem III below), which cannot be performed globally in the whole space.

**4. Local transformation theorem for 3D Ising model**

As pointed out in ref. [45], the existence of the topological problem in the 3D Ising model means the existence of a transformation according to the topological theory [15-17]. Therefore, using approaches of B. Kaufman [18] (see Theorem II above), topology theory (e.g. see L.H. Kauffman [15-17]) and gauge field theory [19,33] we can establish the following theorem for a 3D counterpart of the 2D Ising model of a crystal with an order-disorder transition in its quaternionic formulation by L. Onsager [29].

**Theorem III (Local Transformation Theorem)**

*A local transformation can be performed on the sub-transfer matrices $Sub(V^{(\delta)})$ of the 3D Ising model for each δ. The local transformation changes the gauge of the local systems and trivializes the non-trivial topological structure, while it generates the topological phases on the eigenvectors of the 3D Ising model.*

**Proof**

Following Theorem I, the sub-transfer matrix $V^{(\delta)}$ can be isolated from each other and following Theorem II, the sub-transfer matrix $V^{(\delta)}$ can be linearized, while it splits to be $2^n$ subspaces for each $V^{(\delta)}$. However, the problem remains unsolved. This is because 1) during the linearization process, we cannot keep the eigenvalue to be +1 or -1 as Kaifman did for the 2D Ising model, and 2) the non-trivial topological states exist in the transfer matrices. Indeed, since the subscripts of the matrices in the sub-transfer matrices indicate the existence of crosses formed by the bonds between different sites j, as shown clearly in different terms of exponential factors $\prod_{j=1}^{n}\exp\{iK''\Gamma_{2j}\Gamma_{2j+2n-1}\}$.

According to the topology theory [15-17], a state of the knot diagram is in analogy to the energetic states of a physical system. There is a way to preserve the state structure as the system is deformed topologically, making the invariant properties of states become topological invariants of the knot or link. The topological evolution of states and the integration over the space of states for a given system are complementary in studying the topology of knots and links. Topologically, there are two choices for smoothing a given crossing (×), and thus there are $2^N$ states of a diagram with $N$ crossings [15-17]. The bracket polynomial, i.e., the state summation, defined for knots and links is an analog of a partition function in discrete statistical mechanics, which can be used to express the partition function for the Potts model (also for the Ising model) for appropriate choices of commuting algebraic variables [15-17]. One could transform from the non-trivial topological basis to the trivial topological basis by a transformation, and vice versa. As long as knots or links exist in a system, as revealed in [45], a matrix representing such a transformation (i.e., a rotation) may always exist intrinsically and spontaneously in the system, no matter how complicated the knots or links are.

Therefore, we have to perform a local transformation $\mathbf{G}(j)$ on the j-th site of the 3D Ising system (for every j). An additional exponential factor is generalized for each j site, which is representive of a rotation in a local plane. After performing on all the sites j in the 3D Ising lattice, this process will trivialize the non-trivial topological structure, which changes the system from a basis with non-trivial topological structure to a basis with trivial topological structure. After performing the local transformation, the term in each sub-space of the sub-transfer matrix $V^{(\delta)}$ becomes:

$$V^{(\delta)\pm\ldots\pm} = \frac{1}{2^n} \prod_{j=1}^{n} \circ (1 \pm W_j) \prod_{j=1}^{n} \circ \exp\{iK^* \cdot \Gamma_{2j-1}\Gamma_{2j}\} \prod_{j=1}^{n} \circ \exp\{iK'\Gamma_{2j}\Gamma_{2j+1}\}$$
$$\prod_{j=1}^{n} \circ \exp\{\pm iK''\Gamma_{2j}\Gamma_{2j+1}\} \prod_{j=1}^{n} \circ \exp\{\pm iK'''\Gamma_{2j}\Gamma_{2j+1}\} \quad (14)$$

The non-commutation of operators in the local transformation process also needs to be treated by the Jordan-von Neumann-Wigner procedure [14] with Jordan algebras (see also Theorem IV: Commutation Theorem). All the multiplications in Eq. (14)

satisfy Jordan algebras $A \circ B = \frac{1}{2}(AB + BA)$ [12,13].

The signs $\pm$ stand for splits of sub-spaces owing to the internal factors $W_j$, and the combination of these $\pm$ signs for $W_j$ results in $2^n$ sub-spaces for each sub-transfer matrix $V^{(\delta)}$. The combination of all the $\pm$ signs for $W_j$ in all the sub-transfer matrices $V^{(\delta)}$ ($\delta = 1,2,\ldots,l$) leads to $2^{nl}$ sub-spaces for the whole system (neglecting the effects of the boundary factors $U'$ and $U''$).. After the local transformation, each piece corresponds to a subspace for a free-fermion problem. Moreover, the sub-transfer matrices $V^{(\delta)}$ of the 3D Ising model can be treated as the product of factors like $e^{\theta \Gamma \Gamma / 2}$, being a rotation in a local plane. Among $2^{nl}$ sub-spaces produced by the direct product of all the sub-transfer matrices, the largest eigenvalue is found in the only piece of sub-space. It is reasonable to choose the piece of sub-space with positive signs for all the exponential factors for every j, which contributes to the largest eigenvalue of the system. Actually, we can re-arrange again the sequence of these sub-transfer matrices $V^{(\delta)\pm\ldots\pm}$ and the unit matrices, to collect the basic elements of all the sub-transfer matrices in a sub-space (at least, for that with the largest eigenvalue), so that the product in Eq. (14) can be extended to cover from 1 to nl, with trace invariance (we omit the details for this procedure) For example, in a sub-space with all signs positive $(+\ldots+)$. the direct product of all the sub-transfer matrices $V^{(\delta)+\ldots+}$ over all $\delta$ ($\delta = 1,2,\ldots,l$) results in a term of

$$V^{+\ldots+} = \frac{1}{2^{nl}} \prod_{j=1}^{nl} \circ (1+W_j) \prod_{j=1}^{nl} \circ \exp\{iK^* \cdot \Gamma_{2j-1}\Gamma_{2j}\} \prod_{j=1}^{nl} \circ \exp\{iK'\Gamma_{2j}\Gamma_{2j+1}\}$$
$$\prod_{j=1}^{nl} \circ \exp\{iK''\Gamma_{2j}\Gamma_{2j+1}\} \prod_{j=1}^{nl} \circ \exp\{iK'''\Gamma_{2j}\Gamma_{2j+1}\} \quad (15)$$

All the multiplications in Eq. (15) satisfy Jordan algebras [12-14].

We shall discuss in details the local transformation as well as the topological phases as follows:

At the first of the beginning, the eigenvectors of the 3D Ising model can be constructed from those of the 2D Ising model, since the former should include the latter and can be reduced to the 2D one. Therefore, the eigenvectors of the 3D Ising model have the form of $\psi_{3D} = \psi_{2D}^i \vec{i} + \psi_{2D}^j \vec{j}$, which are in the $2^{nl}$-space of the spin representation. When the direct product of k (= nl(o-1)) terms of unit matrices **I** is added to the transfer matrices (see Theorem I), the dimension of the 3D Ising model is extended to (3+1)-dimensions. Meanwhile, the eigenvectors of the system in the spin representation are extended also to higher dimensions so that we will have the eigenvectors in the quaternion form of $\psi_{(3+1)D} = \psi_{2D}^i \vec{i} + \psi_{2D}^j \vec{j} + \psi_{2D}^k \vec{k}$, which are in the $2^{nlo}$-space of the spin representation. .

The transformation is performed locally on every j site in a sub-transfer matrix $V^{(\delta)}$, which goes on for all the sub-transfer matrices $V^{(\delta)}$ to cover all the j sites in the 3D Ising lattice. The situation is very much similar to the Riemann surface, on which the local surface could be treated as a plane for a rotation within the framework of local coordinates. The 3D Ising model as a whole can be treated on a Riemann surface since the non-trivial topological structure exists. However, each sub-transfer matrix is a quasi-2D system so that one can perform a rotation in a plane within the framework of local coordinates. This local transformation could generate a topological phase by a monodromy representation. The local transformation can be mapped to a gauge transformation. The 3D Ising model is a duality of the 3D $Z_2$ Ising lattice gauge theory via the Kramers-Wannier duality transformation [19,33], in which one can perform a gauge transformation G(n) that flips all the spins on links connected to a site n, keeping the invariance of the Action [19]. The gauge transformation changes the spin $A_{\mu;i} = e^{iT_{\mu;i}}$ with $T_{\mu;i} = 0, \pi$ by the operation $T_{\mu;i} \to T'_{\mu;i} = T_{\mu;i} + \Delta_\mu \Lambda_i$, while it generalizes the phase factor on wave functions (or fields) of the system [33]. Here the nearest neighbor difference operator $\Delta_\mu$ is defined by $\Delta_\mu c_i \equiv c_i - c_{i-\bar\mu}$ and $\Lambda_i$ is an arbitrary scalar defined on the sites of the 3D Ising gauge lattice, which is dual to the original 3D Ising model lattice taking on values 0 or $\pi$ on each site [33]. It is understood that because of the duality, the local transformations on each site j (as discussed above) and their combination of the 3D Ising model can be equivalent to the gauge transformation on the site n of the 3D Ising lattice gauge model, vice versa [19,33]. Acting on every sub-transfer matrix of the 3D Ising model locally and successively "opens" the knots represented by these Pauli matrices. Therefore, it transforms the system from a non-trivial topological basis to a trivial basis (see Eq. (28) of [45]), agreeing with the topological theory [15-17]. Such a local transformation (i.e., a rotation) is represented by an additional matrix with the rotation angle $K'''$ as proposed in ref. [44], which exists intrinsically and spontaneously in the system [45]. Meanwhile, performing the local gauge transform on the transfer matrices **V = V₃V₂V₁** generates the topological phases on the eigenvectors of the 3D Ising system with invariance of the Action (Lagrangian), which corresponds to the phase factors of the gauge transformation in the 3D Ising gauge lattice [19,33]. The eigenvectors become $\psi'_{(3+1)D} = w_x \psi^i_{2D} \vec{i} + w_y \psi^j_{2D} \vec{j} + w_z \psi^k_{2D} \vec{k}$. Here, as discussed previously in ref. [45,46], the weight factors $w_x$, $w_y$ and $w_z$ can be written as $w_x = |w_x| e^{i\phi_x}$, $w_y = |w_y| e^{i\phi_y}$ and $w_z = |w_z| e^{i\phi_z}$, respectively. When $|w_x|$, $|w_y|$ and $|w_z|$ are set to be unit, the phases $\phi_x$, $\phi_y$ and $\phi_z$ determine the weight factors $w_x$, $w_y$ and $w_z$, respectively. Correspondingly, in the rotation representation, we will have the formulas for the 2nlo-normalized quaternion eigenvectors with the topological phases in the rotation representation (see Eq. (33) of [44]). The emergence of topological phases is also a direct result of the transformation on a circuit path on the Riemann surface, which represent the global behavior of the system.

∎

**Remark:**

The Theorem III provides a procedure to trivialize the non-trivial topological structure in the 3D Ising model and to take into account the contributions of non-trivial topological structures to the partition function, the free energy and the thermodynamic properties of the 3D Ising model. The topological phase generalized in the 3D Ising model originates from the non-trivial behavior of a complex projective Hilbert space, which is equivalent to the gauge potential in the parametric space, and is determined by the geometry as a whole of the path.

It should be noted that the linearization process and the local transformation can be performed either simultaneously or subsequently. Clearly, after the local transformation, all the factors in the transfer matrix **V'** contains only quadratic terms of Γ-matrices. Namely, the transfer matrix **V** with higher order terms of Γ-matrices is trivialized. Therefore, the transfer matrix **V'** consists of only fermionic Gaussian terms and becomes commutative. We can apply the generalized Yang-Baxter equation or the so-called tetrahedron equation to commute all the factors of the transfer matrix

**V'**(U) with **V'**(V) [45]. Hence the system becomes integrable.

We shall fix the rotation angle $K'''$ and the phase factors $\phi_x$, $\phi_y$ and $\phi_z$ in Sections 5 and 6, respectively.

**5. Commutation theorem for 3D Ising model**

We can prove the following theorem:

**Theorem IV (Commutation Theorem)**

*The non-commutative behaviors of the operators in the linearization and local transformation processes of the 3D Ising model can be dealt with to be commutative in the framework of the Jordan-von Neumann-Wigner procedure with Jordan algebras, by performing a time average of t systems of the 3D Ising models.*

**Proof**

The algebraic approach to quantum mechanics can be based on the Jordan algebras [12,13]. The landmarks paper [14] by P. Jordan, von Neumann and Wigner provides the mathematical basis of quantum theory and generalizes it on the base of the Jordan algebras. Thanks to the main theorem of [14] and earlier results of P. Jordan [12,13], one realizes the importance of Jordan algebras and of structures generated from the quaternionic structure in comparison with those generated from the octonionic structure. N. Bohr's concept of complementarity [2] found a very impressive representation in the mathematical scheme of quantum mechanics: Starting from a simple 3D theory of material waves in Schrödinger's sense, one could quantize this theory and so come back to the Hilbert space of quantum mechanics. The complete equivalence of the particle and wave pictures of the quantum theory had found its rigorous basis. The Heisenberg's Uncertainty principle, Pauli exclusion principle and the Bose statistics also achieved their proper place in the quantum theory [8].

As pointed out in ref. [20-22], one of the present authors (ZDZ) gave quaternion-based 3D (quantum) models of order-disorder transition for simple orthorhombic Ising lattices, based on Jordan-von Neumann-Wigner procedure [14]. The Zhang's situation is related to the quaternionic sequence of Jordan algebras implied by the fundamental paper of Jordan, von Neumann, and Wigner [14]. Such quaternionic sequence of P. Jordan algebras can be used to bring Zhang's model to a more elegant form [20-22]. The (3+1)-dimensional Ising model in a certain order, being the evolution of the system in space-time, can be transformed into a 3D Ising model, vice versa. In the following process, we show in details how to discuss the (3+1)-dimensional Ising model using the Jordan algebra, with commutative operators/matrices, and transform the results to the 3D Ising model.

Two matrices **A** and **B**, which are in Jordan algebra [12-14]., one has:

$$A \circ B = \frac{1}{2}(AB + BA) \tag{16}$$

N matrices $A_1$, $A_2$, $A_3$, …$A_n$, which are in Jordan algebra, one has the following generalization:

$$(A_1 \circ A_2)A_3 A_4 ... A_n = \frac{1}{2}(A_1 A_2 + A_2 A_1)A_3 A_4 ... A_n \tag{17}$$

……

$$A_1 \circ A_2 \circ A_3 \circ A_4 \circ A_5 ... \circ A_n$$
$$= \frac{1}{2^{n-1}}((A_1 A_2 A_3 A_4 A_5 ... A_{n-1} A_n) + (A_2 A_1 A_3 A_4 A_5 ... A_{n-1} A_n)$$
$$+ (A_3 A_1 A_2 A_4 A_5 ... A_{n-1} A_n) + (A_3 A_2 A_1 A_4 A_5 ... A_{n-1} A_n)$$
$$+ (A_4 A_1 A_2 A_3 A_5 ... A_{n-1} A_n) + (A_4 A_2 A_1 A_3 A_5 ... A_{n-1} A_n)$$
$$+ (A_4 A_3 A_1 A_2 A_5 ... A_{n-1} A_n) + (A_4 A_3 A_2 A_1 A_5 ... A_{n-1} A_n) \quad (18)$$
$$+ ......$$
$$+ (A_{n-3} A_{n-2} A_{n-1} A_n A_{n-4} ... A_2 A_1) + (A_{n-3} A_{n-2} A_n A_{n-1} A_{n-4} ... A_2 A_1)$$
$$+ (A_{n-3} A_{n-1} A_n A_{n-2} A_{n-4} ... A_2 A_1) + (A_{n-3} A_n A_{n-1} A_{n-2} A_{n-4} ... A_2 A_1)$$
$$+ (A_{n-2} A_{n-1} A_n A_{n-3} A_{n-4} ... A_2 A_1) + (A_{n-2} A_n A_{n-1} A_{n-3} A_{n-4} ... A_2 A_1)$$
$$+ (A_{n-1} A_n A_{n-2} A_{n-3} A_{n-4} ... A_2 A_1) + (A_n A_{n-1} A_{n-2} A_{n-3} A_{n-4} ... A_2 A_1))$$

Besides the method introduced in Theorem I, there are two other ways to introduce the additional dimension: 1) to simply sum t copies of the 3D Ising model and then make an average by dividing it by t, which does not change the physical property of the system, but does not solve the problem since the internal factors are not removed out and the transfer matrices are still not commutative; 2) to stack t copies (i.e., t slices) of the 3D Ising model in the (3+1)-dimensional space-time, with evolution of the positions of the factors (or operators) in the transfer matrices. The results obtained by these two paths are equivalent, however, only does the latter procedure work for finding the desired solution. Clearly, from Eq. (18), introducing the t slices of the 3D Ising model in the (3+1)-dimensions would result in the multiplication of Jordan algebra in transfer matrices (and/or sub-transfer matrices) of the 3D Ising model, see Eqs. (12)-(15), which are commutative.

In statistical mechanics, the ergodic hypothesis states that the time average of a quantity over the time evolution of a specific microstate equals the average of the same quantity over some statistical ensembles of microstates at fixed time [47]. Acceptance of the ergodic hypothesis implies that the use of a statistical ensemble is justified provided the time necessary for an efficient sweep of the ensemble by any of its microstates is short enough compared with the time of measurement of the physical quantity of interests [7]. In the 3D Ising model, the existence of internal factors $W_j$ in the transfer matrices makes serious challenge to the validation of the ergodic hypothesis [47]. One cannot neglect the time average as in other models of statistical mechanics, since it appears to deal with the non-commutative problem caused by these internal factors $W_j$ (standing for the global effect, or the non-trivial topological effect). The time necessary for the time averaging must be infinite, being comparable with or even much longer than the time of measurement of the physical quantity of interest. Therefore, it is necessary to perform the 4-fold integral on the partition function for the 3D Ising model, as derived in [44] to meet the requirement of taking the time average. In this way, the time is maintained in the framework of quantum statistical mechanics for the time averaging physical quantities in an equilibrium system. This framework provides the opportunity for constructing the quaternionic sequence of Jordan algebras [12,13], so that the Jordan-von Neumann-Wigner procedure [14] can be employed to deal with the 3D Ising model, as illustrated in [44] also see [45]. The unit quaternions can be thought of as a choice of a group structure on the 3-sphere $S^3$ that gives the group Spin(3), which is isomorphic to SU(2) and also to the universal cover of SO(3). The quaternionic bases found in ref. [44] for the 3D Ising model are complexified quaternionic bases, on the 3-sphere $S^3$ that gives the group Spin(3), which is isomorphic to SU(2) and also to the universal cover of SO(3). Therefore, the quaternion basis constructed in ref. [44] for the 3D Ising model represents naturally a rotation in the 4D space (a (3 + 1) – dimensional space-time). This procedure is related closely with well-developed theories, for example, complexified quaternion [4], quaternionic quantum mechanics [1,6,26], and quaternion and special relativity [3].

∎

**Remark:**

The Theorems I-III and IV are consistent and complementary in the sense of introducing the additional dimension. In Theorems I-III, the 3D Ising model is expanded to be (3 + 1) dimensional by adding k terms of unit matrices in the direct product of the original transfer matrices and by extending the eigenvectors of the system in the spin representation also to higher dimensions. In Theorem IV, the 3D Ising model is expanded to be (3 + 1) dimensional by performing a time average of t systems of the 3D Ising models and by constructing a framework of quaternion basis in consistent with the Jordan-von Neumann-Wigner procedure and Jordan algebras. Theorem IV solves the non-commutation problem of operators in Theorems II and III, and guarantees the validity of Theorems II and III.

On the other hand, as pointed out in ref. [45], tetrahedron equation, or generalized Yang-Baxter equation, was introduced to deal with the 3D Ising models [11,33,42,43], which preserve the commutativity of layer-to-layer transfer matrices constructed from the weight functions,. analogous to the Yang-Baxter equation for the 2D Ising model, The 3D statistical system can be treated as a 2D system with a composite weight [35] The trick is that, projecting the cubic lattice along the third direction results in a quadratic lattice with the effective Boltzmann weight [35]. One can deduce global properties like the commutativity of transfer matrices of the 3D Ising model, by imposing the tetrahedron equation on the local statistical weights of the model. These local symmetry relations can be applied to derive the global properties of the model, since one can associate local statistical weights at all the intersections of the lattice, in such a way that the tetrahedron and inverse relations are satisfied everywhere on the lattice, and the transformations will leave the partition function unchanged [11]. In one word, the commutativity of transfer matrices and the integrability of the 3D Ising model are guaranteed by tetrahedron equations, or generalized Yang-Baxter equation.

**6. Rotation angle of local transformation for 3D Ising model**

According to the procedures above (Theorems I - IV), a local transformation is performed to smooth the non-trivial knot/link structures of the 3D Ising model, which transforms the basis from the non-trivial topological one to the trivial one, generalizing the topological phases on the eigenvectors. Then the transfer matrices consist of quadratic expressions in the $\Gamma$-matrices only, which correspond to the spinor representation of the Lie algebra for rotations. It indicates that we can follow the Onsager-Kaufman procedure [18,29] to diagonalize the transfer matrices and derive a formula for the partition function of the 3D Ising model (Eq. (49) of ref. [44] or Eq. (24) of ref. [45]), which is similar to that for the 2D Ising model, but within the quaternionic framework with appearance of topological phases (see Eqs. (21)-(32) and Eqs. (34)-(48) of ref. [44]). The aim of this section is to figure out the rotation angle $K'''$ for local transformation (or the scaling factor $\lambda$). The topological phases will be discussed in details in the next section.

The local transformation is a spin representative of a rotation in a local plane, which looks like to move the nearest neighboring spin along the third dimension to the nearest neighboring position along the second dimension. This also corresponds to the interchange of two spins. It is interesting to figure out the rotation angle $K'''$ for the local transformation. At first, we should point out that $K'''$ takes the same value for all the sub-transfer matrices $V^{(\delta)}$ (and all the j sites). We have $K''' = \lambda K'' \propto K''$. For the symmetry of the transfer matrices $V_2$ and $V_3$ (the high-order terms of $\Gamma$ matrices can appear in either $V_2$ or $V_3$ up to our choose for the second and third dimensions), we have also $K''' = \lambda' K' \propto K'$. Therefore, we have $K''' \propto K'K''$. In order to figure completely the rotation angle $K'''$ for the local transformation, the star-triangle relationship is employed for Curie temperature, which is the solution of generalized Yang-Baxter equations in the continuous limit [45]. For the 3D simple orthorhombic Ising lattice with $K, K', K''$ as its interactions along three dimensions, we can find a dual simple orthorhombic lattice with three interactions $K*, K'*, K''*$ [44,45], following the Kramers-Wannier duality transformation [18,19,29,33]. Each edge of the new lattice is dual to each face of the unit cell of the old one, which connects the center of the unit cell of the old one. The duality relationship exists between these interactions $K, K', K''$ and $K*, K'*, K''*$. It is noticed that $K*$ appears in the transfer matrix $V_1$, while $K', K''$ appear in $V_2$ and $V_3$. To ensure the commutativity of the transfer matrices and the integrability of the 3D Ising model, the transfer matrices should satisfy the generalized Yang-Baxter equation (so-called tetrahedron equation). It is known that the star-triangle relation is the solution of the Yang-Baxter equation in the continuous limit. At Curie temperature, the correlation length approaches infinite, so that the critical region of the 3D Ising model can be treated as a continuous one. Therefore, the relation for the

Curie temperature of the 3D Ising model should be one of the following star-triangle relations [44,45]: $KK^* = K'K'^* = K''K''^* = KK'+K'K''+K''K$. Since we have assumed K is the largest one among the three interactions $K, K', K''$, the reasonable relation with physical significance is $KK^* = KK'+K'K''+K''K$, i.e., $K^* = K'+K''+(K'K''/K)$. On the other hand, according to Theorems I - IV proved above, we can arrive the partition function (Eq. (49) of ref. [44] or Eq. (24) of ref. [45]) and the condition for the Curie temperature (Eqs. (50) and (51) of ref. [44]), we have $K^* = K'+K''+K'''$. By comparison, we see that the rotation angle for the local transformation is $K''' = K'K''/K$, while the scaling factor $\lambda = 1 + K'/K$.

We arrive at the following statement:

*For the local transformation of the 3D simple orthorhombic Ising lattice, we have the rotation angle $K''' = K'K''/K$.*

## 7. Topological phases for 3D Ising model

The 3D spin Ising model can be mapped into the 3D $Z_2$ Ising gauge lattice model by the Kramers-Wannier duality transformation [19,33]. It is easier to understand the emergence of phase factors on the eigenvectors with the framework of the Ising lattice gauge theory. The Action of the Ising lattice gauge theory is associated with the product of four spins $A_\nu$ located on links which border the plaquette [33]. Defining $A_{\mu;i} = e^{iT_{\mu;i}}$, the local gauge transformation is described as $T_{\mu;i} \to T'_{\mu;i} = T_{\mu;i} + \Delta_\mu \Lambda_i$ and, $\psi \to \psi' = \exp(i\Lambda_i)\psi$ [19,33]. In the Ising lattice gauge theory, the phase factors are generalized on the eigenvectors (wave functions), which should be 0 or $\pi$ [33].

Such a process should correspond to the emergence of phase factors for a local transformation in the original 3D Ising model. We have to fix the values of these phase factors. For the 3D Ising model, the procedure of introducing the direct products of unit matrices in the transfer matrices (Theorem I) extends the dimension of the 3D system to be a (3+1)D one. The topological phases are generalized on the quaternion eigenvectors in the $2^{nlo}$-space of the spin representation, so that we have $\psi_{(3+1)D} \rightarrow \psi'_{(3+1)D} = w_x \psi^i_{2D} \vec{i} + w_y \psi^j_{2D} \vec{j} + w_z \psi^k_{2D} \vec{k}$ where the weight factors $w_x$, $w_y$ and $w_z$ are determined by the phases $\phi_x$, $\phi_y$ and $\phi_z$, respectively. It is nature to fix one of the three topological phases $\phi_x$, $\phi_y$, and $\phi_z$, without loss of generality, say $\phi_x$, which corresponds to the invariance of the properties of the 2D Ising model (since the 3D Ising model consists of many 2D Ising models, which should be reduced to a 2D Ising model in the 2D limit). That is, $\phi_x = 0$ or $2\pi$. Then we have to fix the phases $\phi_y$ and $\phi_z$, which are generalized by the local transformations in the 3D Ising model. The combination of the two phases should equal to the phase generalized by the gauge transformation in the 3D Ising gauge lattice model. Namely, we have $\phi_y + \phi_z = \pi$. From the symmetry of the system, $\phi_y = \phi_z$. Therefore, $\phi_y = \pi/2$ and $\phi_z = \pi/2$.

We obtain the following result:

*The topological phases $\phi_x$, $\phi_y$, and $\phi_z$ generalized on the eigenvectors of the 3D Ising model by the local transformation are equal to 0 (or $2\pi$), $\pi/2$ and $\pi/2$, respectively.*

It is interesting to inspect further the physical significance of the topological phases in the 3D Ising model. It was pointed out in ref. [44,45] that the topological phases in the 3D Ising model are analogous to those of non-Abelian anyons in fractional quantum Hall systems, which originate from interchanging many-body interacting particles (spins) [27]. However, it is well known that in 3D systems, usually, a process wrapping one particle around another is topologically equivalent to a process in which none of the particles move at all. The wave function should be left unchanged by two such interchanges of particles, with only two possibilities, bosons/fermions with symmetric/antisymmetric wave functions (changed by a ± sign under a single interchange). It is important to understand why spins in the 3D Ising model can have the statistical angle ϕ equal to π/2 (unequal to 0 and π). In the 3D Ising model, the interchange of two spins is actually performed in a 2D plane, since it is a local transformation for a spin and its neighboring spin along the third dimension. The latter spin can be treated as (n+1)-th spin in the plane, so that the interchange of spins is confined in 2D. So, as a 2D system does not necessarily come back to the same state after a nontrivial winding involved in the trajectory of two particles (or spins) which are interchanged twice in a clockwise manner, because it can result in a nontrivial phase $e^{2i\phi}$. The 2n·l·o-normalized eigenvectors for the 3D Ising model is described by quaternion ones with a scalar part and a 3D vector part, while those of the 2D Ising model are in the form of a scalar part and a 1D vector. The interacting spins in the 3D Ising lattice always force themselves to be confined in one of many 2D planes and then move to another, so that the partition function of the 3D Ising model has the feature of the 2D Ising model, but with the non-trivial topologic phases. This is intrinsic requirement of non-trivial topological structures of the 3D Ising system. In addition, in the physics of gauge theories, Wilson lines correspond essentially to the space-time trajectory of a charged particle [37,38]. The particles represented by Wilson lines in the Chern-Simons theory have fractional statistics. Under a change of framing, the expectation values of Wilson lines are multiplied by a phase $\exp(2\pi i h_a)$, where $h_a$ is the conformal weight of the field. A twist of a Wilson line is equivalent to a phase, while a braiding of two Wilson lines from a trivalent vertex is also equivalent to a phase. For the procedure to smooth the knots/crossings, in the 3D Ising model, it is natural that a phase appears in the functions of the bases of the Hilbert space The transformation of a system between different (space-time) frames can bring the gauge potential (or phase factors).

In summary, by utilizing some basic facts of the direct product and the trace, we prove Theorem I which provides a chance to deal with the sub-transfer matrices in the quasi-2D limit, and to overcome the difficulties (such as Non-local, Non-Gaussian and non-commutative) of the 3D Ising problem. By utilizing the same character of the internal factor $W_j$ and the boundary factor U and by following the Kaufman's procedure for 2D Ising model, we perform a linearization process on the non-linear terms of the transfer matrices of the 3D Ising model (Theorem II). By introducing a local transformation, we transfer the 3D Ising model from a non-trivial topological basis to a trivial topological basis, while taking into account of the contribution of the non-trivial topological structures (Theorem III). By utilizing Jordan algebras in the framework of the Jordan-von Neumann-Wigner procedure and by performing a time average, we succeed in dealing with the non-commutation of the operators during the processes of linearization and local transformation (Theorem IV). By fixing the rotation angle for the local transformation and the phase factors, we realize the desired solution. This work provides a mathematical base for the Zhang's two conjectures. The four Theorems proved in this work develop an approach for a topological quantum statistical mechanics, in which the contributions of non-trivial topological structures are dealt with, a time average is considered and the Jordan algebras is utilized in the framework of the Jordan-von Neumann-Wigner procedure. This algebra approach is consistent with the topological approach developed in a recent work [36]. In future work, it will be of interest to inspect the origins of the relation $K''' = K'K''/K$ and the phase factors $\phi_x$, $\phi_y$ and $\phi_z$, from the views of topological and geometrical aspects.

**Acknowledgements**

This work has been supported by the National Natural Science Foundation of China under grant numbers 51331006 and 51590883, by The State key Project of Research and Development of China (No.2017YFA0206302), and by the key project of Chinese Academy of Science under grant number KJZD-EW-M05-3. ZDZ acknowledges Prof. J.H.H. Perk for helpful discussion on properties of the transfer matrices, Prof.Julian Ławrynowicz for discussion on linearization process. ZDZ also is grateful to Fei Yang for understanding, encouragement, support and discussion.

**Appendix A: Transfer matrices of the 3D Ising model**

For the 3D Ising model, from the Hamiltonian (1), one can derive the partition function, which sums up all the possible configurations of the system. Following the Onsager- Kaufman process [18,29], the first transfer matrix $V_1$ is described as

[5,24,44,45]:

$$V_1 = a \otimes a \otimes ... \otimes a. \tag{A1}$$

$$a = \begin{pmatrix} e^{\beta J} & e^{-\beta J} \\ e^{-\beta J} & e^{\beta J} \end{pmatrix} = I \cdot e^{\beta J} + C \cdot e^{-\beta J} = (2\sinh 2K)^{1/2} e^{K^*C} \tag{A2}$$

In Eq. (A1), nl factors of a matrices appear in the product, and therefore the transfer matrix $V_1$ is a $2^{nl}$-dimensional matrix. In this paper, we follow the notation of Onsager-Kaufman-Zhang for Pauli matrices [18,29,44,45]: $s'' = \begin{bmatrix} 0 & -1 \\ 1 & 0 \end{bmatrix}$ (= $i\sigma_2$), $s' = \begin{bmatrix} 1 & 0 \\ 0 & -1 \end{bmatrix}$ (= $\sigma_3$), $C = \begin{bmatrix} 0 & 1 \\ 1 & 0 \end{bmatrix}$ (= $\sigma_1$), $I = \begin{bmatrix} 1 & 0 \\ 0 & 1 \end{bmatrix}$ and we have the relation $\tanh K^* = e^{-2K}$. We introduce $K = \beta J$, $K' = \beta J'$ and $K'' = \beta J''$ with $\beta = 1/(k_B T)$. Therefore,

$$V_1 = (2\sinh 2K)^{nl/2} e^{K^*C} \otimes e^{K^*C} \otimes ... \otimes e^{K^*C} \tag{A3}$$

We can prove the following formula:

$$\begin{aligned} e^{K^*C} \otimes e^{K^*C} &= (I \cdot \cosh K^* + C \sinh K^*) \otimes (I \cdot \cosh K^* + C \sinh K^*) \\ &= [(I \otimes I)\cosh K^* + (C \otimes I)\sinh K^*] \cdot [(I \otimes I)\cosh K^* + (I \otimes C)\sinh K^*] \\ &= e^{K^*(C \otimes I)} \cdot e^{K^*(I \otimes C)} = e^{K^*C_1} \cdot e^{K^*C_2} \end{aligned} \tag{A4}$$

There is equivalent between the two hands of the above formulas, the direct product of the exponential factors of Pauli matrices C and the product of the exponential factors of matrices $C_j = I \otimes I \otimes ... \otimes I \otimes C \otimes I \otimes ... \otimes I$ ($C_j$ is the direct product of Pauli matrix C and unit matrices, where nl factors appear in each product $C_j$ while C appears in the j-th place). This equation can be generalized so that the transfer matrix $V_1$ can be transformed from the direct product of the exponential factors of Pauli matrices to the product of the exponential factors of the direct product of Pauli matrices:

$$V_1 = (2\sinh 2K)^{nl/2} e^{K*C} \otimes e^{K*C} \otimes ... \otimes e^{K*C}$$

$$= (2\sinh 2K)^{nl/2} \prod_{j=1}^{nl} \exp(K*C_j) = (2\sinh 2K)^{nl/2} \prod_{j=1}^{nl} \exp\{iK* \cdot \Gamma_{2j-1}\Gamma_{2j}\}$$

(A5)

Note that Eq. (A5) is consistence with the result represented in Eqs. (2) and (5), although the transfer matrix $V_1$ is redefined so as to remove the scalar coefficient $(2\sinh 2K)^{nl/2}$.

The similar situation is for the second transfer matrix $V_2$ [5,18,24,29,44,45].

$$V_2 = (b \otimes b \otimes ... \otimes b)(\bullet b \otimes b \otimes ... \otimes b),$$  (A6)

$$b = \begin{pmatrix} e^{K'} & 0 & 0 & 0 \\ 0 & e^{-K'} & 0 & 0 \\ 0 & 0 & e^{-K'} & 0 \\ 0 & 0 & 0 & e^{K'} \end{pmatrix} = \cosh K' \begin{pmatrix} 1 & 0 & 0 & 0 \\ 0 & 1 & 0 & 0 \\ 0 & 0 & 1 & 0 \\ 0 & 0 & 0 & 1 \end{pmatrix} + \sinh K' \begin{pmatrix} 1 & 0 & 0 & 0 \\ 0 & -1 & 0 & 0 \\ 0 & 0 & -1 & 0 \\ 0 & 0 & 0 & 1 \end{pmatrix}$$

$$= (I \otimes I)\cosh K' + (s' \otimes I)(I \otimes s')\sinh K' = e^{K's'_1 s'_2}$$  (A7)

We can prove the following formula:

$$b \otimes b = e^{K's'_1 s'_2} \otimes e^{K's'_1 s'_2}$$
$$= ((I \otimes I)\cdot \cosh K' + s'_1 s'_2 \sinh K') \otimes ((I \otimes I)\cdot \cosh K' + s'_1 s'_2 \sinh K')$$
$$= [(I \otimes I \otimes I \otimes I)\cosh K' + (s'_1 s'_2 \otimes I \otimes I)\sinh K'] \cdot$$
$$[(I \otimes I \otimes I \otimes I)\cosh K' + (I \otimes I \otimes s'_1 s'_2)\sinh K']$$
$$= e^{K'(s'_1 s'_2 \otimes I \otimes I)} \cdot e^{K'(I \otimes I \otimes s'_1 s'_2)} = e^{K's'_1 s'_2} \cdot e^{K's'_3 s'_4}$$  (A8)

with $s'_j = I \otimes I \otimes ... \otimes I \otimes s' \otimes I \otimes ... \otimes I$ ($s'_j$ is the direct product of Pauli matrix $s'$ and unit matrices, where $s'$ is in the j-th place of the direct product). Therefore, the direct product of nl/2 terms of b matrices is written as: $b \otimes b \otimes ... \otimes b = e^{K's'_1 s'_2} \cdot e^{K's'_3 s'_4} ... \cdot e^{K's'_{nl-1} s'_{nl}}$. The number of b matrices involved in the direct product above for $V_2$ is half of that of a matrices for the transfer matrix $V_1$, since b is 4×4 matrix while a is 2×2 matrix. This takes into account of only half of exponential terms of the transfer matrix $V_2$. It is seen from Eq. (A6) that the transfer matrix $V_2$ is obtained by the multiply of $(b \otimes b \otimes ... \otimes b)$ and $(\bullet b \otimes b \otimes ... \otimes b)$. Here we use • to denote that there is a shift of a unit matrix in the second term of the direct product of b matrices, which represents another half of the exponential factors for each element of the transfer matrix $V_2$. We have $\bullet b \otimes b \otimes ... \otimes b == e^{K's'_2 s'_3} \cdot e^{K's'_4 s'_5} ... \cdot e^{K's'_{nl} s'_{nl+1}}$. For a system of the finite size, one has to consider the periodic boundary condition, which may generalize some boundary factors $U'_s$ (see Eq. (12) in ref. [45].). The boundar factor $U'_s$ are the same as those in the 2D Ising model. However, in the case of thermodynamic limit (n → ∞, l → ∞), the boundary factors can be neglected, since the surface to volume ration vanishes for an infinite system according to the Bogoliubov inequality. According to Eq. (A6) and discussions above, we obtain:

$$V_2 = \prod_{j=1}^{nl} \exp(K' s'_j s'_{j+1}) \tag{A9}$$

Note that Eq. (A9) is equivalent to Eq. (4).

Similarly, we have the third transfer matrix $V_3$ [24,44,45].,

$$V_3 = (c \otimes c \otimes ... \otimes c)(\bullet c \otimes c \otimes ... \otimes c)...(\bullet ... \bullet c \otimes c \otimes ... \otimes c)$$
$$= \prod_{j=1}^{nl} \exp(K'' s'_j s'_{j+n}) \tag{A10}$$

Note that Eq. (A10) is equivalent to Eq. (3), in which the internal factors appear the exponential factors for every j. The internal factors $W_{r+1,s}$ are represented also in Eq.(11) and (14a) of ref. [45].

We have used the following relations:

$$c \otimes c = e^{K''s'_1 s'_{1+n}} \otimes e^{K''s'_1 s'_{1+n}}$$
$$= ((I \otimes I ... \otimes I) \cdot \cosh K'' + s'_1 s'_{1+n} \sinh K'') \otimes ((I \otimes I ... \otimes I) \cdot \cosh K'' + s'_1 s'_{1+n} \sinh K'')$$
$$= [(I \otimes I \otimes ... \otimes I \otimes I) \cosh K'' + (s'_1 s'_{1+n} \otimes I \otimes ... \otimes I) \sinh K''] \cdot$$
$$[(I \otimes I \otimes ... \otimes I \otimes I) \cosh K'' + (I \otimes I \otimes ... \otimes I \otimes s'_1 s'_{1+n}) \sinh K'']$$
$$= e^{K''(s'_1 s'_{1+n} \otimes I \otimes ... \otimes I)} \cdot e^{K''(I \otimes I \otimes ... \otimes I \otimes s'_1 s'_{1+n})} = e^{K''s'_1 s'_{1+n}} \cdot e^{K''s'_{2+n} s'_{2+2n}}$$

(A11)

The c matrix is $2^{n+1} \times 2^{n+1}$ matrix. The direct products of l terms of c matrices result in

$c \otimes c \otimes ... \otimes c = e^{K"s'_1 s'_{1+n}} \cdot e^{K"s'_{2+n} s'_{2+2n}} ... e^{K"s'_{l+n(l-1)} s'_{l+nl}}$. It is seen that the above formula only consists of l term of exponential factors, and also one has to consider the periodic boundary condition, since the size of the matrix is $2^{(n+1)l} \times 2^{(n+1)l}$. In order to take into account all the nl exponential terms in the transfer matrix $V_3$, one has to perform the multiply for n times of the direct product of l terms of c matrices and considere the shifts of unit matrices as denoted again by •, which generalize the boundary factors U"$_r$ (see Eq. (11) in ref. [45]). Once again, the extra boundary factors U"$_r$ (together with the internal factor $W_{r+1,l}$ in Eq. (14b) of ref. [45]) can be negligible in the thermodynamical limit.

**Remark:**

As revealed above (and also in, [49]), the interaction between the most neighboring spins along the third dimension looks like an interaction between two spins located far from each other, via a chain of n spins in the plane (actually, via nm spins in the plane if one considers the periodic condition already used along the first dimension,). The reason is that the third transfer matrix $V_3$ must follow the sequence of the spin (σ) variables arranged and fixed already in the first two transfer matrices **V₁** and **V₂**. According to this fixed order, although the interaction between a spin and one of its neighboring spins along the third dimension is the nearest neighboring one, its effect is correlated with the states of nm other spins in the plane. It is equivalent effectively to a long-range and many-body interaction in which (nm+1) spins are involved. The same effect happens for every interaction along the third dimension in the 3D Ising model. Therefore, the non-local correlation and the global effect indeed exist in the 3D Ising system. From another angle of points of view, all the spins in the 3D Ising model are entangled.

**Note added after published in Advances in Applied Clifford Algebras 29 (2019) 12:**

Although the Ising model looks like to be fully locally defined in the original Ising spin variable language, the set of all allowed states contribute to partition function and free energy in a way of all spins entangled. It is clearly seen from above results that the non-locality shows up in the alternative Clifford algebra description, defined through auxiliary fermionic Γ-operators, the new method developed in this work deals with the non-locality in this Clifford algebra space. It is understood that the non-locality exists not only in the space of description of Γ-operators, but also in the space of all the Ising spin states. Although it is not evidently and clearly seen, the descriptions in the two different spaces are connected by a series of equalities. Therefore, the non-locality indeed exists in the original Ising spin variable language. It is clear now that although the Ising model with only nearest-neighboring interactions seems look like local in the original spin language, one cannot neglect the existence of the non-trivial topological effect in the system. The situation is very much similar to Aharonov-Bohm (A-B) effect in which in the language of magnetic field, no field, but in the language of potential, the A-B effect exists. However, one cannot say that the A-B effect does not exist in the system (even in the language of magnetic field). The A-B effect revealed in the language of potential is observable quantity of topological phases with physical significance. Similarly, the topological effect (i.e. non-locality) revealed in the language of Γ-operators for the 3D Ising model is also observable quantity with physical significance. One cannot neglect its existence even in the original spin variable language. In other words, we think that the original spin variable language is not a good representation for studying the 3D Ising model, as the language of magnetic field for the A-B effect.

**Appendix B: Some facts of the direct product of matrices**

We recall some facts for the direct product of matrices.

We choose $A, A' \in M_n(C), B, B' \in M_m(C), \ldots\ldots Z, Z' \in M_m(C)$, with

$$A = \begin{pmatrix} a_{11} & a_{12} \\ a_{21} & a_{22} \end{pmatrix}, \quad A' = \begin{pmatrix} a'_{11} & a'_{12} \\ a'_{21} & a'_{22} \end{pmatrix}, \quad B = \begin{pmatrix} b_{11} & b_{12} \\ b_{21} & b_{22} \end{pmatrix}, \quad B' = \begin{pmatrix} b'_{11} & b'_{12} \\ b'_{21} & b'_{22} \end{pmatrix}, \ldots\ldots$$

$$Z = \begin{pmatrix} z_{11} & z_{12} \\ z_{21} & z_{22} \end{pmatrix}, \quad Z' = \begin{pmatrix} z'_{11} & z'_{12} \\ z'_{21} & z'_{22} \end{pmatrix}$$

The direct product is defined as:

$$A \otimes B = \begin{bmatrix} a_{11}b_{11} & a_{11}b_{12} & a_{12}b_{11} & a_{12}b_{12} \\ a_{11}b_{21} & a_{11}b_{22} & a_{12}b_{21} & a_{12}b_{22} \\ a_{21}b_{11} & a_{21}b_{12} & a_{22}b_{11} & a_{22}b_{12} \\ a_{21}b_{21} & a_{21}b_{22} & a_{22}b_{21} & a_{22}b_{22} \end{bmatrix}$$

1) We can easily prove that $(A \otimes B) \cdot (A' \otimes B') = (A \cdot A') \otimes (B \cdot B')$ and its general form

$$(A \otimes B \otimes \ldots \otimes Z) \cdot (A' \otimes B' \otimes \ldots \otimes Z') = (A \cdot A') \otimes (B \cdot B') \otimes \ldots \otimes (Z \cdot Z').$$

2) We recall a property of the direct product of matrices:

$$Trace[1 \otimes A \otimes B \otimes ... \otimes Z] = 2 \times Trace[A \otimes B \otimes ... \otimes Z]$$

$$Trace[1 \otimes ... \otimes 1 \otimes A \otimes B \otimes ... \otimes Z] = 2^K \times Trace[A \otimes B \otimes ... \otimes Z]$$

3) We also mention another property of the direct product of matrices:

$Trace[A \otimes B] = Trace[B \otimes A] = (A_{11} + A_{22}) \cdot (B_{11} + B_{22})$. This indicates that the circling (or permutation) of the elements in the direct product of matrices does not affect the trace. Namely, $Trace[A \otimes B \otimes ... \otimes Z] = Trace[Z \otimes ... \otimes B \otimes A]$

4) We may use . $A \otimes B \otimes C = A \otimes (B \otimes C)$ .

*to Norman H. March on the occasion of his 90th birthday*", G. G. N. Angilella and C. Amovilli, editors (New York : Springer, 2018)